# AEDNet: Adaptive Edge-Deleting Network For Subgraph Matching

Zixun Lan [a], Ye Ma [b], Limin Yu [c], Linglong Yuan [d], Fei Ma [a, *]

[a] *Department of Applied Mathematics, School of Mathematics and Physics, Xi'an Jiaotong-Liverpool University, Suzhou, China*
[b] *Department of Financial Mathematics, School of Mathematics and Physics, Xi'an Jiaotong-Liverpool University, Suzhou, China*
[c] *Department of Electrical and Electronic Engineering, Xi'an Jiaotong-Liverpool University, Suzhou, China*
[d] *Department of Mathematical Sciences, The University of Liverpool, United Kingdom*

zixun.lan19@student.xjtlu.edu.cn, {ye.ma, limin.yu, fei.ma}@xjtlu.edu.cn, linglong.yuan@liverpool.ac.uk

## Abstract

Subgraph matching is to find all subgraphs in a data graph that are isomorphic to an existing query graph. Subgraph matching is an NP-hard problem, yet has found its applications in many areas. Many learning-based methods have been proposed for graph matching, whereas few have been designed for subgraph matching. The subgraph matching problem is generally more challenging, mainly due to the different sizes between the two graphs, resulting in considerable large space of solutions. Also the extra edges existing in the data graph connecting to the matched nodes may lead to two matched nodes of two graphs having different adjacency structures and often being identified as distinct objects. Due to the extra edges, the existing learning based methods often fail to generate sufficiently similar node-level embeddings for matched nodes. This study proposes a novel Adaptive Edge-Deleting Network (AEDNet) for subgraph matching. The proposed method is trained in an end-to-end fashion. In AEDNet, a novel sample-wise adaptive edge-deleting mechanism removes extra edges to ensure consistency of adjacency structure of matched nodes, while a unidirectional cross-propagation mechanism ensures consistency of features of matched nodes. We applied the proposed method on six datasets with graph sizes varying from 20 to 2300. Our evaluations on six open datasets demonstrate that the proposed AEDNet outperforms six state-of-the-arts and is much faster than the exact methods on large graphs.

## 1. Introduction

Subgraph matching, which is to find all subgraphs in a data graph $G$ that are isomorphic to a query graph $Q$, has found its applications in various fields. In information retrieval, a query process of searching for research papers is a subgraph matching task [1]. In computer vision, images and objects can be converted into graphs through preset rules, which enables object recognition to be treated as a subgraph matching problem [2–4]. In natural language processing, words in the corpus can be treated as nodes and relationships between words as edges, so paraphrases are equivalent to subgraph matching [5]. In chemoinformatics, template-based methods take the template's reaction centre and product molecules as the query and data graph respectively. The first step of applying the template on the product molecule becomes the process of subgraph matching [6]. In bioinformatics, a considerable amount of biological data can be naturally represented by graphs, thus a query operation is also a subgraph matching task [7,8].

Subgraph isomorphism is a generalization of the graph isomorphism problem. Compared with the graph matching problem, the subgraph matching problem is generally more challenging, mainly due to the different sizes between the two graphs, resulting in the considerable large space of solutions [9,10]. Subgraph isomorphism can be divided into induced and non-induced subgraph isomorphism problems [11]. In this study, we focus on the induced subgraph isomorphism problem, as the experimental data used in this study contains only induced subgraph matches. However, theoretically, our method can also be used for non-induced subgraph isomorphism problem.

There are exact methods for subgraph matching, such as Ullman's [12], VF2 [10], Ceci [13], FilM [14], VF3 [15], etc, but they are difficult to meet real-world situations. Exact methods incur a significant computational burden when the number of nodes is large, which is often the case in many real applications. In addition, the graphs in reality often have noises, causing difficulties to find subgraphs isomorphic to query graph and further resulting in no matching after a long time searching. To solve the subgraph matching problem within a reasonable time and with a noisy back-ground, one has to look for fast inexact methods that can tolerate the existence of noise.

* Corresponding author.
  *E-mail addresses:* zixun.lan19@xjtlu.edu.cn (Z. Lan), ye.ma@xjtlu.edu.cn (Y. Ma), limin.yu@xjtlu.edu.cn (L. Yu), Linglong.Yuan@liverpool.ac.uk (L. Yuan), fei.ma@xjtlu.edu.cn (F. Ma).

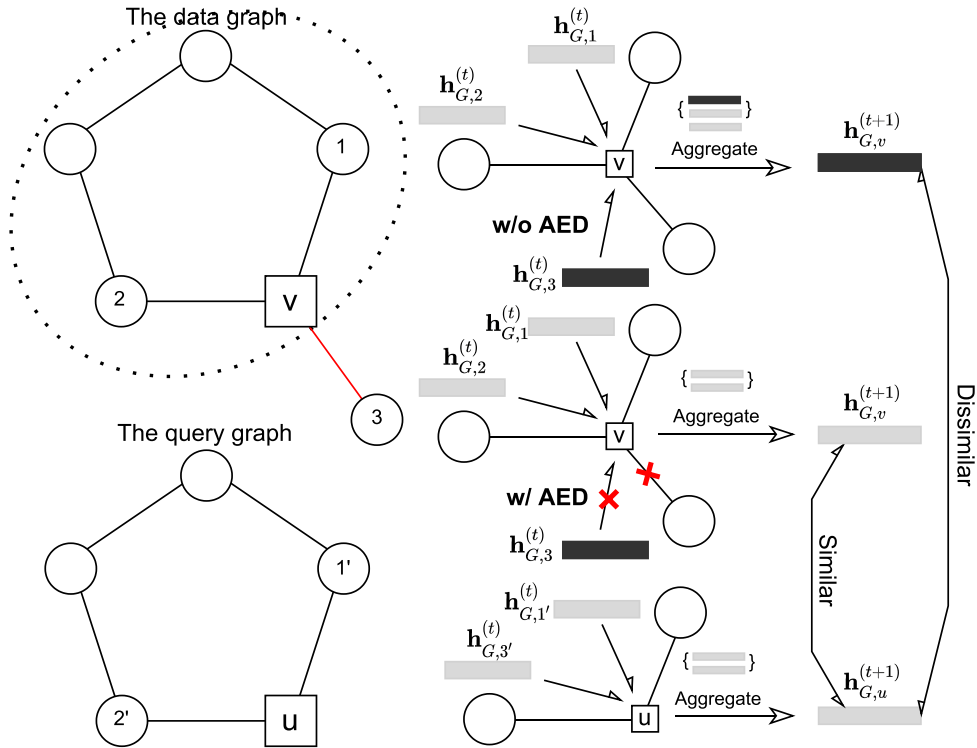

**Fig. 1.** Subgraph Matching: the subgraph of the data graph in the ellipse is isomorphic to the query graph. The red edge is an extra edge. AED represents the Adaptive Edge-Deleting mechanism. Node $u$ in the query graph and node $v$ in the data graph is a pair of matched nodes. The **h**, i.e. grey and black rectangular blocks, represent node-level embeddings respectively. Due to the extra edge, $\mathbf{h}_{G,v}^{(t+1)}$ does not match $\mathbf{h}_{G,u}^{(t+1)}$. Removing the extra edge by AED can make the embeddings of node $u$ and node $v$ similar enough. This figure shows part of the whole model. Our model is end-to-end. The Input of the model is two graphs and the output of the model is the predicted matching matrix. (For interpretation of the references to color in this figure legend, the reader is referred to the web version of this article.)

Various approximate algorithms have been proposed to solve subgraph matching, including G-Finder [16], Saga [17], PG-N [18] and so on. These methods often work in a fast but heuristic way in which they select appropriate seed nodes and then expand to their neighbours continually with preset rules. In contrast, rarely have learning-based methods been proposed, especially for subgraph matching. Compared with more time-consuming and power-intensive traditional methods which are mostly performed on CPU and have more logic operations [19], learning-based approaches learn matching relationships from data and are normally very efficient during the inferencing due to highly parallel computation and full usage of GPUs. During the training, the parameters are learned by minimizing the difference between the ground-truth and the predicted matching; during the testing, unseen pairs (data graph and query graph) can be fed into the models for fast approximation of matching relationships.

Many learning-based methods have been proposed for a similar problem, graph matching, such as Zanfir and Sminchisescu [20], Wang et al. [21], Ma et al. [22], Wu et al. [23], Xu et al. [24], Guo et al. [25], Vento [26]. However, very few learning-based methods have been designed for subgraph matching, except for Sub-GMN [27] and NeuralMatch [28] to our best knowledge.

The extra edges lead to the subgraph matching problem being more challenging than graph matching from the perspective of deep learning and graph representation learning [29]. In more detail, a matched node $v$ in the data graph may be connected with extra edges (the red line in Fig. 1), but no corresponding edges are connecting to the matched node $u$ in the query graph, where $u$ and $v$ are a pair of matched nodes. It means that a pair of matched nodes from the data graph and the query graph respectively are distinct objects from the node-level perspective because the adjacency structures of the matched nodes are inconsistent, even if their node features are identical.

The existing learning-based subgraph matching approaches usually follow the learning-based graph matching methods. One shared spatial graph convolution network computes the node-level embedding of each node in two graphs, and then the embeddings of two matched nodes are expected to be close to each other by minimizing the distance or maximizing the similarity. However, as mentioned before, the adjacency structures of the matched nodes in the data graph and the query graph are inconsistent due to the extra edges (Fig. 1), so two matched nodes in subgraph matching can be easily identified as different objects. This direct approach in the subgraph matching is difficult to make the embeddings of two matched nodes to be similar enough. It is valid for graph matching since two matched nodes are identical objects, but this approach fails to force two different objects to output the same representation for subgraph matching.

This paper is inspired by the disharmony between inputting distinct objects and outputting similar embeddings. Two matched nodes with different adjacency structures are represented as distinct embeddings through one shared spatial graph convolution. However, in subgraph matching we always expect the embeddings of two matched nodes to be close to each other. This study proposes a novel sample-wise adaptive edge-deleting mechanism to ensure that two matched nodes have close adjacent structures to predict the node-to-node matching relationships in a more reasonable way. A unidirectional cross-propagation mechanism is also used to ensure that the node features of the corresponding nodes are similar enough. We propose an Adaptive Edge-Deleting Network (AEDNet) for the approximate subgraph matching problem in this study which combines the above two mechanisms. The model is trained in an end-to-end fashion (see Fig. 2). Since the extra edges are known during the training, the model can be trained to adaptively 'delete' edges by reducing the weight of the extra edges in the data graph according to the corresponding input query graphs. Through AEDNet, two matched nodes with different adjacent structures could eventually be mapped together with close representations. Our contributions are summarized as below:

- We identify the difficulty with the most current learning-based subgraph matching methods that is to force two different objects to output the same representation vectors through one vanilla shared graph network.
- We propose a novel sample-wise adaptive edge-deleting mechanism to ensure that two matched nodes are consistent in terms of structures. We showed both mathematically (Theorem 1) and experimentally the effectiveness of the proposed sample-wise adaptive edge-deleting mechanism.
- Our evaluations on six open datasets demonstrate that the proposed AEDNet outperforms the comparison state-of-the-arts and is faster than exact methods.

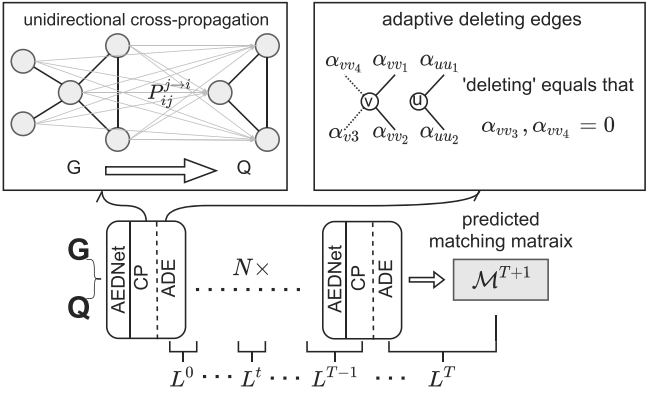

**Fig. 2.** Here $v \in G$ and $u \in Q$ are corresponding nodes, dotted lines represent extra edges, $\alpha_{ij}$ is the normalized attention coefficient. In this example, $A^G_{nd(v)} = \{\alpha_{vv_1}, \alpha_{vv_2}\}$, $A^G_{d(v)} = \{\alpha_{v_3}, \alpha_{vv_4}\}$, $A^Q_{(u)} = \{\alpha_{uu_1}, \alpha_{uu_2}\}$, where 'd' is the abbreviation of 'delete' and 'nd' is the abbreviation of 'no-delete'. The sample-wise adaptive edge-deleting mechanism in the model makes the sum of all elements in $A^G_{d(v)}$ tend to 0, which leads to $A^G_{nd(v)} = A^Q_{(u)}$.

## 2. Related work

### 2.1. Learning graph matching

Many learning-based graph matching methods are developed for specific scenarios. DLGM [20] and PCA-GM [21] match images by predicting corresponding nodes after images are converted into graphs. In the field of knowledge graph, RDGCN [23] and GMNN [24] are used for entity alignment, where RDGCN uses a dual relation counterpart to improve its expressive ability. In action recognition, NGMN [25] combines few-shot learning with GCN [29] to match graphs representing actions. The above methods use GNNs to represent nodes and then directly force node-level embeddings of corresponding nodes close to each other. It is reasonable since corresponding nodes are objects with the same adjacency structures and identical node features in the graph matching task. However, these methods are not suitable for generic graph matching or subgraph matching because the sizes of graphs are kept below 20 under these specific scenarios. In subgraph matching there is often an extreme imbalance of sizes between data graphs and query graphs.

Graph similarity calculation is an important graph-level task. Methods such as GMN [30] and INFMCS [31] target at this task. Their internal mechanisms relate to and contribute to the graph matching task, especially the cross-attention in GMN and cross-interaction in Simgnn, namely the cross-propagation mechanism.

### 2.2. Learning subgraph matching

Following the assumption of learning-based graph matching methods, Sub-GMN [27] directly forces node-level embeddings of two matched nodes to be close to each other. It is inappropriate, as the corresponding nodes of the data graph may have extra edges, making the corresponding nodes be different objects. Sub-GMN aims to make the corresponding node-level embeddings similar by combining GCN [29] and NTN [32], which inevitably leads to poor performance since it maps two different objects to close points in representation space forcibly.

NeuralMatch [28] first decomposes the graph pairs into a large number of k-order Breadth First Search (BFS) small graphs anchored at each node in data graphs and query graphs. Subsequently, it assumes that the decomposed k-order BFS subgraph anchored at a node of the query graph is a subgraph of the k-order BFS subgraph anchored at the corresponding node of the data graph. After that, it trains GNNs to generate graph-level embedding of one graph at the bottom left of that of another graph in the embedding space according to the order embedding [33] if this graph is a subgraph of another graph. Even if this assumption is valid on ground-truth corresponding nodes, it may still result in mismatches, since the k-order BFS subgraph anchored at one unmatched node in the data graph may contain a subgraph isomorphic to the k-order BFS subgraph anchored at one node in the query graph.

### 2.3. Learning-free methods on subgraph matching

There are two main categories of subgraph matching approaches. There are a large number of subgraph matching models which can be classified as backtracking search based algorithms, such as Kim et al. [34], Jüttner and Madarasi [35]. These models obtain matches on the query graph by the filtering, ordering and enumeration operation. Another category is the join-based methods, such as Yang et al. [36], Dahm et al. [37]. The join-based approaches decompose the query graph into nodes and edges. Then, they repeatedly perform join operations to integrate matched ones into query. In addition, there are some subgraph matching methods [18,38] which are based on constrained programming. There are also some algorithms [16,39] based the indexing-enumeration. They build an index on the data graph and answer all queries with the help of the index.

## 3. Problem definition and preliminary

### 3.1. Subgraph matching and matching matrix

A graph is denoted as a tuple $\{V, E, F^n, F^e\}$, where $V$ is the node set, $E$ is the edge set. $F^n$ and $F^e$ are feature functions which map each node and each edge to feature vectors respectively. In this paper, $Q = \{V_Q, E_Q, F^n_Q, F^e_Q\}$ represents a query graph, and $G = \{V_G, E_G, F^n_G, F^e_G\}$ stands for a data graph.

**Definition 1.** The subgraph matching [9] is an injection function $m : V_Q \to V_G$, which satisfies: (1) $\forall u \in V_Q, m(u) \in V_G$ and $F^n(u) = F^n(m(u))$; (2) $\forall (u_a, u_b) \in E_Q, (m(u_a), m(u_b)) \in E_G$ and $F^e(u_a, u_b) = F^e(m(u_a), m(u_b))$. There may be multiple mappings from $V_Q$ to $V_G$. We use $SM(Q, G) = \{m_1, m_2, \ldots, m_k\}$ for the set of all mappings. The subgraph matching problem is to find the set of all mappings $SM(Q, G)$ for a given pair of graphs.

**Definition 2.** A Ground-Truth Matching Matrix [40] gives the true node-to-node matching relationship. The Matching Matrix $M \in R^{|Q| \times |G|}$ is defined as:

$$M_{ij} = \begin{cases} 1 & m_n(i) = j \\ 0 & m_n(i) \neq j, n = 1, 2, \ldots, k \end{cases} \quad (1)$$
$$M = [M_{ij}]_{|Q| \times |G|},$$

where $i$, $j$ are the $i$th row and the $j$th column of $M$, corresponding to the $i$th and the $j$th nodes of $Q$ and $G$ respectively. $M_{ij}$ is the element of $i$th row and $j$th column of $M$. $m_n(i) = j$ means the $i$th node of $Q$ maps to the $j$th node of $G$. $|Q|$ and $|G|$ are the sizes of the query graph $Q$ and the data graph $G$ respectively. Matching Matrix $M$ contains all the node-to-node matching relationships, so at least one element in each row of the Matching Matrix is 1.

Notably, in this paper, we focus on top-1 subgraph matching. We found that although multiple subgraphs of $G$ could be isomorphic to $Q$, most of the nodes in these subgraphs are repeated, and only a few nodes are changing. It shows that finding the top-1 subgraph is equivalent to finding the most critical nodes of other isomorphic subgraphs. On the other hand, neural-based methods are generally divided into two steps to solve combinatorial optimization problems [19]. They firstly predict a probability transition matrix and then select the solution from the probability transition matrix through specific heuristic rules. This paper focuses on a predicted probability transition matrix (Matching Matrix) rather than following heuristic rules. Intuitively, a better-predicted probability transition matrix often leads to a better solution (top-1 subgraph).

*3.2. Graph attention network*

Graph Attention Network (GAT) [41] integrates attention mechanism [42] with spatial graph convolution to obtain expressive power.

Given $\mathbf{H}^{(t)} = [\mathbf{h}_1^{(t)}; \mathbf{h}_2^{(t)}; \cdots; \mathbf{h}_n^{(t)}] \in R^{n \times d}$ is the node embedding matrix at the $t$th layer, where $\mathbf{h}_i^{(t)} \in R^{1 \times d}$ is the node-level embedding for node $i$ of the graph and is also the $i$th row of $\mathbf{H}^{(t)}$, $d$ is the dimension of node-level embedding and $n$ is the number of nodes. GAT injects the graph structure into the attention mechanism by performing masked attention, namely it only computes $\alpha_{ij}$ for nodes $j \in \mathcal{N}_i$, where $\mathcal{N}_i$ is the first-order neighbors of node $i$ in the graph:

$$e_{ij} = \mathbf{a} \cdot \left[ \mathbf{h}_i^{(t)} \mathbf{W} \| \mathbf{h}_j^{(t)} \mathbf{W} \right]^T,$$
$$\alpha_{ij} = \frac{\exp(\theta(e_{ij}))}{\sum_{k \in \mathcal{N}_i} \exp(\theta(e_{ik}))}, \quad (2)$$

where $e_{ij} \in R$ and $\alpha_{ij} \in R$ are a non-normalized attention coefficient and a normalized attention coefficient representing the weight of message aggregated from node $j$ to node $i$ respectively, and $\|$ is the concatenation operation. Besides, $\mathbf{a} \in R^{1 \times 2d'}$ and $\mathbf{W} \in R^{d \times d'}$ are learnable parameters ($d'$ is a hyperparameter), $\theta$ is a LeakyReLU nonlinearity with negative input slope $\alpha = 0.2$.

GAT employs multi-head attention to stabilize the learning process of self-attention, similar to Transformer [43]. If there are $K$ heads, $K$ independent attention mechanisms execute the Eq. (2), and then their features are concatenated:

$$\mathbf{h}_i^{(t+1)} = \|_{k=1}^{K} \sigma \left( \sum_{j \in \mathcal{N}_i} \alpha_{ij}^k \mathbf{h}_j^{(t)} \mathbf{W}^k \right) \quad (3)$$
$$= Aggr(\{\alpha_{ij}^k, \mathbf{h}_j^{(t)} | j \in \mathcal{N}_i\})$$

where $\|$ represents concatenation, $\alpha_{ij}^k$ are normalized attention coefficients computed by the $k$th learnable $\mathbf{a}^k \in R^{1 \times 2d'}$ and $\mathbf{W}^k \in R^{d \times d'}$ following Eq. (2), $Aggr()$ is the final aggregation function.

## 4. Model: adaptive edge-deleting network

We propose an Adaptive Edge-Deleting Network inspired by the disharmony mentioned above. The unidirectional cross-propagation mechanism and the sample-wise adaptive edge-deleting mechanism ensure that the node features and node adjacency structures of matched nodes are similar. Fig. 2 is an example of AEDNet.

*4.1. Unidirectional cross-propagation*

Unlike the previous graph matching task [21,30], we use unidirectional cross-propagation instead of bidirectional mode:

$$P_{ij}^{|G| \to |Q|, (t)} = \mathcal{M}_{ij}^{(t)} = \frac{\exp\left(s_h\left(\mathbf{h}_i^{Q,(t)}, \mathbf{h}_j^{G,(t)}\right) \times \tau_*^{-1}\right)}{\sum_{j'} \exp\left(s_h\left(\mathbf{h}_i^{Q,(t)}, \mathbf{h}_{j'}^{G,(t)}\right) \times \tau_*^{-1}\right)},$$
$$\mathbf{v}_i^{Q,(t)} = \sum_j P_{ij}^{|G| \to |Q|, (t)} \mathbf{h}_j^{(t)}, \quad (4)$$
$$\forall i \in V_Q, j \in V_G,$$

here $\mathbf{h}_i^{Q,(t)} \in R^{1 \times d}$, $\mathbf{h}_j^{G,(t)} \in R^{1 \times d}$ are node-level embeddings at $t$th layer for query graph $Q$ and data graph $G$ respectively, $s_h$ is a vector space similarity metric, like Euclidean or cosine similarity. $P^{|G| \to |Q|, (t)} \in R^{|Q| \times |G|}$ is the propagation matrix from data graph $G$ to query graph $Q$ at the $t$th layer, and $\mathcal{M}^{(t)} \in R^{|Q| \times |G|}$ is the predicted matching matrix at the $t$th layer. Notably, their calculation processes are identical. We use $\mathbf{N}_Q^{(t)} = [\mathbf{v}_1^{Q,(t)}; \mathbf{v}_2^{Q,(t)}; \cdots; \mathbf{v}_{|Q|}^{Q,(t)}] \in R^{|Q| \times d}$ as the cross-information matrix of the query graph $Q$ aggregated from the data graph $G$ at the $t$th layer and $\mathbf{v}_i^{Q,(t)} \in R^{1 \times d}$ is the cross-information of one node of query graph aggregated from data graph. In order to discretize propagation matrix $P^{|G| \to |Q|, (t)}$ for specific situations, we add a learnable parameter $\tau_* \in (0, 1]$.

The purpose of cross-propagation is to make the cross-information of the corresponding node features similar enough when propagation matrix $P^{|G| \to |Q|}$ is close enough to the Ground-Truth Matching Matrix $M$. The hidden representations of the corresponding nodes at the first layer are nonidentical in subgraph matching due to extra edges. So, using bidirectional-propagation inevitably leads to different representations of the corresponding nodes.

We design a loss function $L_{\mathcal{M}}^{(t)}$ to meet the above purpose and to make the predicted matching matrix $\mathcal{M}^{(t)}$ to be closer to the Ground-Truth:

$$\varrho_i = \sum_{j' \in \mathcal{J}_i} \mathcal{M}_{ij'}^{(t)}, \mathcal{J}_i = \{j'|M_{ij'} = 1\}$$
$$\rho_i = \sum_{j' \in J_i} \mathcal{M}_{ij'}^{(t)}, J_i = \{j'|M_{ij'} = 0\} \quad (5)$$
$$\forall i \in V_Q, j' \in V_G$$
$$L_{\mathcal{M}}^{(t)} = \frac{1}{|Q|} \sum_{i \in V_Q} \|\mathcal{V}_i - 1\|_2, \mathcal{V}_i = \varrho_i - \rho_i$$

where $M \in R^{|Q| \times |G|}$ is the ground-truth matching matrix, $\varrho_i \in R$ ($\rho_i \in R$) is the sum of normalized similarity between node $i$ and nodes $j'$ (un)matched by node $i$ according to the ground-truth, and $\mathcal{V}_i$ represents the difference between $\varrho_i$ and $\rho_i$. We expect $\mathcal{V}_i$ to be equal to one due to normalization, which leads the embeddings of corresponding nodes to be more similar than embeddings of these unmatched nodes.

*4.2. Sample-wise adaptive edge-deleting*

Given $v \in G$, $u \in Q$ are corresponding nodes and $h_v^{G,(t)} = \mathbf{v}_u^{Q,(t)}$ ($P^{|G| \to |Q|} = M$), we use $\mathbf{N}_Q^{(t)}$ instead of $\mathbf{H}_Q^{(t)}$ to compute the non-normalized attention coefficient $e_{ij}$ and the normalized attention coefficient $\alpha_{ij}$ at the $t$th layer defined in Eq. (2), and then obtain

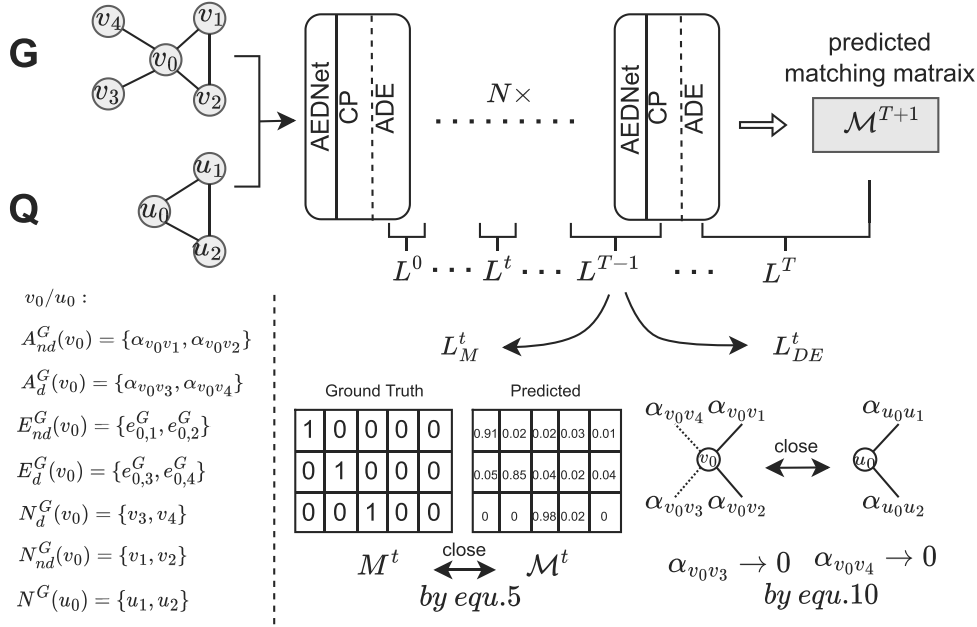

**Fig. 3.** Left: shows an example for set of *N*, *A* and *E* in Eq. (6). Right: shows $L_{\mathcal{M}}^{(t+1)}$ and $L_{DE}^{(t)}$ briefly.

weight sets:

$$\begin{aligned}
\mathbf{N}_{nd(v)}^{G} &= \{j_v = m(j_u) | (u, j_u) \in E_Q, (v, m(j_u)) \in E_G\},\\
\mathbf{N}_{d(v)}^{G} &= N_v / \mathbf{N}_{nd(v)}^{G},\\
\mathbf{E}_{nd(v)}^{G,(t)} &= \{e_{vj}^{(t)} | j \in \mathbf{N}_{nd(v)}^{G}\}, \mathbf{A}_{nd(v)}^{G,(t)} = \{\alpha_{vj}^{(t)} | j \in \mathbf{N}_{nd(v)}^{G}\},\\
\mathbf{E}_{d(v)}^{G,(t)} &= \{e_{vj}^{(t)} | j \in \mathbf{N}_{d(v)}^{G}\}, \mathbf{A}_{d(v)}^{G,(t)} = \{\alpha_{vj}^{(t)} | j \in \mathbf{N}_{d(v)}^{G}\},\\
\mathbf{E}_{(u)}^{Q,(t)} &= \{e_{uj}^{(t)} | j \in N_{(u)}\}, \mathbf{A}_{(u)}^{Q,(t)} = \{\alpha_{uj}^{(t)} | j \in N_{(u)}\},
\end{aligned} \quad (6)$$

where superscript $G$ and $Q$ denote the data and query graph respectively. $v \in G$ and $u \in Q$ are a pair of corresponding nodes. 'd' and 'nd' in the subscript are the abbreviation of 'delete' and 'no-delete'. $d(v)$ denotes the nodes or edges linking to or equal to the extra edges to be 'deleted', and $nd(v)$ denotes the nodes or edges linking to or equal to the edges not to be 'deleted'. $\mathbf{N}_{nd(v)}^{G}$ and $\mathbf{N}_{d(v)}^{G}$ are the sets of first-order neighbors not associated and associated with the extra edges of nodes $v$ respectively, $\mathbf{E}$ and $\mathbf{A}$ are the corresponding weight sets of $e_{ij}$ and $\alpha_{ij}$, $N_{()}$ represents the first-order neighbors, and $E_G$, $E_Q$ are edge sets for the data graph and the query graph respectively. An example is shown in Fig. 3.

When $P^{|G| \to |Q|} = M$ $(h_v^{G,(t)} = v_u^{Q,(t)})$, $\mathbf{E}_{(u)}^{Q,(t)} = \mathbf{E}_{nd(v)}^{G,(t)}$ can be calculated by using $\mathbf{N}_Q^{(t)}$ instead of $\mathbf{H}_Q^{(t)}$ in Eq. (2). Obtaining the same embedding by Eq. (3) for node $u$ and $v$ needs not only their node features to be the same, but also their adjacency structures to be consistent, namely $A_{(u)}^{Q,(t)} = A_{nd(v)}^{G,(t)}$. However,

$$\begin{aligned}
\mathbf{A}_{(u)}^{Q,(t)} &= softmax(\mathbf{E}_{(u)}^{Q,(t)})\\
&= softmax(\mathbf{E}_{nd(v)}^{G,(t)})\\
&\neq softmax(\mathbf{E}_{nd(v)}^{G,(t)} \cup \mathbf{E}_{d(v)}^{G,(t)}) / \mathbf{A}_{d(v)}^{G,(t)}\\
&= \mathbf{A}_{nd(v)}^{G,(t)}
\end{aligned} \quad (7)$$

due to the extra edges. Actually, $A_{(u)}^{Q,(t)} = A_{nd(v)}^{G,(t)}$ only when $\sum_{\alpha \in A_{d(v)}^{G,(t)}} \alpha = 0$. During the training we know which edge's weight should be 0 in the training set and we use these information as the supervision information.

Randomly initialized learnable query vector **a** in Eq. (2) is challenging to satisfy the above conditions because all samples share **a** at each layer. Thus, we need a new query vector $q^{(t)}$ that can contain the query graph information sample-wisely. In other words, each sample has a specific $q^{(t)}$ at one layer:

$$q^{(t)} = \text{MLP}(\text{Pooling}(\mathbf{H}_Q^{(t)})), \quad (8)$$

here $q^{(t)} \in R^{1 \times 2d'}$ is a sample-wise query vector in the attention mechanism, and the shape of $q^{(t)}$ is the same as **a**. $\mathbf{H}_Q^{(t)}$ is the node embedding matrix for query graph $Q$ at the $t$th layer, Pooling is the pooling operation (e.g., sum, mean, max, min, attention), MLP is a multilayer perceptron that transforms the dimension of the graph-level embedding of the query graph $Q$ to the dimension of **a**.

We replace **a** in Eq. (2) with $q^{(t)}$ in order to sample-wise adaptively delete extra edges:

$$\begin{aligned}
\alpha_{ij}^{G,(t)} &= \frac{\exp\left(\theta\left(q^{(t)} \cdot \left[\mathbf{h}_i^{G,(t)} \mathbf{W} \| \mathbf{h}_j^{G,(t)} \mathbf{W}\right]^T\right)\right)}{\sum_{k \in \mathcal{N}_i} \exp\left(\theta\left(q^{(t)} \cdot \left[\mathbf{h}_i^{G,(t)} \mathbf{W} \| \mathbf{h}_k^{G,(t)} \mathbf{W}\right]^T\right)\right)},\\
\alpha_{ij}^{Q,(t)} &= \frac{\exp\left(\theta\left(q^{(t)} \cdot \left[\mathbf{v}_i^{Q,(t)} \mathbf{W} \| \mathbf{v}_j^{Q,(t)} \mathbf{W}\right]^T\right)\right)}{\sum_{k \in \mathcal{N}_i} \exp\left(\theta\left(q^{(t)} \cdot \left[\mathbf{v}_i^{Q,(t)} \mathbf{W} \| \mathbf{v}_k^{Q,(t)} \mathbf{W}\right]^T\right)\right)}
\end{aligned} \quad (9)$$

where $\alpha_{ij}^{G,(t)}$ and $\alpha_{ij}^{Q,(t)}$ are normalized attention coefficients for the data graph $G$ and the query graph $Q$ respectively. Notably, we replace $\mathbf{h}_i^{Q,(t)}$ with $\mathbf{v}_i^{Q,(t)}$ based on the unidirectional cross-propagation mechanism. In order to obtain the same adjacent structures between the corresponding nodes, namely all elements in $A_{d(v)}^{G,(t)}$ are expected to be 0, we design a loss function $L_{DE}^{(t)}$ for adaptively deleting edges at the $t$th layer:

$$\begin{aligned}
\mathscr{V}_{(v)}^{(t)} &= \sum_{\alpha \in \mathbf{A}_{nd(v)}^{G,(t)}} \alpha - \sum_{\alpha \in \mathbf{A}_{d(v)}^{G,(t)}} \alpha,\\
L_{DE}^{(t)} &= \frac{1}{|Q|} \sum_{v \in \Delta} \|\mathscr{V}_{(v)}^{(t)} - 1\|_2,\\
\Delta &= \{j | m(i) = j, i \in V_Q\},
\end{aligned} \quad (10)$$

where $\mathbf{A}_{nd(v)}^{G,(t)}$ and $\mathbf{A}_{d(v)}^{G,(t)}$ can be found according to the ground-truth label during the training. We expect $\mathscr{V}_{(v)}^{(t)}$ to be equal to one, which leads the sum of weights of the extra edges of the corresponding nodes in the data graph to be close to zero. The following theorem proves its effectiveness mathematically.

**Theorem 1.** Given $\Delta$, $\mathbf{H}_G^{(t)}$ and $q^t$ as defined above, there exists $\mathbf{H}_G^{(t)}$ and $q^t$ such that $F(\mathbf{H}_G^{(t)}, q^t) = \sum_{v \in \Delta} \sum_{\alpha \in \mathbf{A}_{d(v)}^{G,(t)}} \alpha = 0$.

**Proof.**

$$\begin{aligned}F(\mathbf{H}_G^{(t)}, q^{(t)}) &= \sum_{v \in \Delta} \sum_{\alpha \in \mathbf{A}_{d(v)}^{G,(t)}} \alpha \\ &= \sum_{v \in \Delta} \sum_{j \in N_{d(v)}^G} \frac{e^{e_{vj}^{(t)}}}{\sum_{i \in N_{(v)}} e^{e_{vi}^{(t)}}} \\ &= \sum_{v \in \Delta} \frac{\sum_{j \in N_{d(v)}^G} e^{e_{vj}^{(t)}}}{\sum_{i \in N_{(v)}} e^{e_{vi}^{(t)}}} \end{aligned} \quad (11)$$

where $e_{vi}$ is the non-normalized attention coefficient.

$F(\mathbf{H}_G^{(t)}, q^t) = 0$ if and only if $\sum_{v \in \Delta} \sum_{j \in N_{d(v)}^G} e^{e_{vj}^{(t)}} = 0$. To show that the above equation can be met, we will show for $\forall \epsilon > 0$, $\sum_{v \in \Delta} \sum_{j \in N_{d(v)}^G} e^{e_{vj}^{(t)}} < \epsilon$ can be reached.

Let $e_{vi}^{(t)'} = \max\{e_{vi}^{(t)} | v \in \Delta, i \in N_{d(v)}^G\}$, we want to show that:

$$\sum_{v \in \Delta} \sum_{j \in N_{d(v)}^G} e^{e_{vj}^{(t)}} \leq |\Delta| \cdot |N_{d(v)}^G| \cdot e^{e_{vi}^{(t)'}} < \epsilon \quad (12)$$

For simplicity, let $\mathrm{V} = \mathbf{h}_v^{(t)} \mathbf{W} \| \mathbf{h}_i^{(t)} \mathbf{W}^T$. To show

$$q^{(t)} \cdot \mathrm{V} = e_{vi}^{(t)'} < \frac{\epsilon}{|\Delta| \cdot |N_{d(v)}^G|},$$

let $\mathbf{h}_v = \mathbf{h}_i^{(t)} = \{\frac{1}{d}\}^{1 \times d}$, $\mathbf{W} = \{1\}^{d \times d'}$ and $q^{(t)} = [\mathbf{q}_1^{(t)}, \mathbf{q}_2^{(t)}, \ldots, \mathbf{q}_{2d'}^{(t)}]$ ($\mathbf{q}_i^{(t)} = d, i = 1, \ldots, 2d'$), so $\mathrm{V} = \mathbf{h}_v^{(t)} \mathbf{W} \| \mathbf{h}_i^{(t)} \mathbf{W}^T = \{1\}^{2d' \times 1}$.

We have,

$$\begin{aligned} q^{(t)} \cdot \mathrm{V} &= \mathbf{q}_1^{(t)} + \sum_{i=2}^{2d'} \mathbf{q}_i^{(t)} \times 1 \\ &= \mathbf{q}_1^{(t)} + (2d' - 1) \times d \end{aligned} \quad (13)$$

If $\mathbf{q}_1^{(t)} < \frac{\epsilon}{|\Delta| \cdot |N_{d(v)}^G|} - (2d' - 1) \times d$, we will have

$$q^{(t)} \cdot \mathrm{V} < \frac{\epsilon}{|\Delta| \cdot |N_{d(v)}^G|},$$

and hence $\sum_{v \in \Delta} \sum_{j \in N_{d(v)}^G} e^{e_{vj}^{(t)}} = 0$.

Therefore, if we take $\mathbf{q}_1^{(t)} = \frac{\epsilon}{|\Delta| \cdot |N_{d(v)}^G|} - (2d' - 1) \times d - 1$,

$$F(\mathbf{H}_G^{(t)}, q^{(t)}) = \sum_{v \in \Delta} \sum_{\alpha \in \mathbf{A}_{d(v)}^{G,(t)}} \alpha = 0.$$

□

Subsequently, we add the residual connection to Eq. (3) after aggregating multi-headed attention:

$$\begin{aligned} \mathbf{h}_{Gi}^{(t+1)} &= \mathrm{MLP}(Aggr(\{\alpha_{ij}^{G,(t),k}, \mathbf{h}_{Gj}^{(t)} | j \in \mathcal{N}_i\})) + \mathbf{h}_{Gi}^{(t)}, \\ \mathbf{h}_{Qi}^{(t+1)} &= \mathrm{MLP}(Aggr(\{\alpha_{ij}^{Q,(t),k}, \mathbf{v}_{Qi}^{(t)} | j \in \mathcal{N}_i\})) + \mathbf{h}_{Qi}^{(t)}, \end{aligned} \quad (14)$$

where MLP transforms the dimension of the output to the dimension of final output $\mathbf{h}_i^{(t+1)}$ of the last layer.

### 4.3. Training and loss

The overall structure of the model consists of multiple layers of AEDNet. The unidirectional Cross-Propagation mechanism is not included in the first layer since $\mathbf{H}^{(0)}$ does not contain structural information, and there is no $L_{PM}$ in the first layer. On the other hand, after the $T$th layer, the model derives $\mathbf{H}_G^{(T+1)}$ and $\mathbf{H}_Q^{(T+1)}$ is used to calculate $\mathcal{M}^{(T+1)}$, which means that $L_{DE}^{(t)}$ and $L_{\mathcal{M}}^{(t+1)}$ can be paired. We design a total loss $L_{total}$ in order to take into account two mechanisms:

$$\begin{aligned} L^t &= \lambda_1 L_{DE}^{(t)} + (1 - \lambda_1) L_{\mathcal{M}}^{(t+1)}, \\ L_{total} &= \lambda_2 \sum_1^{T-1} L^t + (1 - \lambda_2) L^T \end{aligned} \quad (15)$$

where $L^t$ is the loss at the $t$th layer, $T$ is the number of AEDNet layers, $\lambda_1 \in [0, 1]$ and $\lambda_2 \in [0, 1]$ are hyperparameters that regulates the trade-off between the two components. Fig. 3 shows two losses briefly.

### 4.4. Pseudo-code

The Algorithm 1 describes the forward and backward process of our proposed model. From step 4 to step 5, the AEDNet layer

---

**Algorithm 1** FORWARD and BACKWARD of AEDNet.

**Input**: $\mathbf{H}_Q^{(t)}$ and $\mathbf{H}_G^{(t)}$, the $t$th layer features. $K$, the number of attention heads, $T$, the number of layers
**Parameter**: w = $\{\theta^{k,t}, \theta^t, \mathbf{W}^{k,t}\}$, trained parameters of the AEDNet layer
**Output**: $\mathbf{H}_Q^{(t+1)}$ and $\mathbf{H}_G^{(t+1)}$, the $(t+1)$th layer features. $\mathcal{M}^{(t)}$, the predicted matching matrix at $t$th layer.

1: **FORWARD**($G, Q$; w)
2:  $t \leftarrow 0$
3:  **while** $t <= T$ **do**
4:   // unidirectional cross-propagation Eq. (4).
5:   bulid $\mathcal{M}^{(t)}$ and $v_i^{Q,(t)}$ from $\mathbf{H}_Q^{(t)}$, $\mathbf{H}_G^{(t)}$ by Eq. (4).
6:   // adaptive edge-deleting mechanism Eq. (8, 9, 14).
7:   $k \leftarrow 1$.
8:   **while** $k <= K$ **do**
9:    $q^{(t),k} \leftarrow \mathrm{MLP}_{\theta^{k,t}}(\mathrm{Pooling}(\mathbf{H}_Q^{(t)}))$
10:   compute $\alpha_{ij}^{G,(t),k}, \alpha_{ij}^{Q,(t),k}$ from $\mathbf{N}_Q^{(t)}, \mathbf{H}_G^{(t)}, q^{(t),k}$ by Eq. 9
11:   $k \leftarrow k + 1$
12:  **end while**
13:  $\mathbf{h}_{Gi}^{(t+1)} \leftarrow \mathrm{MLP}_\theta \left( \|_{k=1}^K \sigma \left( \sum_{j \in \mathcal{N}_i} \alpha_{ij}^{G,(t),k} \mathbf{h}_{Gi}^{(t)} \mathbf{W}^{k,t} \right) \right) + \mathbf{h}_{Gi}^{(t)}$
14:  $\mathbf{h}_{Qi}^{(t+1)} \leftarrow \mathrm{MLP}_\theta \left( \|_{k=1}^K \sigma \left( \sum_{j \in \mathcal{N}_i} \alpha_{ij}^{Q,(t),k} \mathbf{v}_{Qi}^{(t)} \mathbf{W}^{k,t} \right) \right) + \mathbf{h}_{Qi}^{(t)}$
15: **end while**
16: **return** $\{\mathbf{H}_Q^{(t+1)}, \mathbf{H}_G^{(t+1)}, \mathcal{M}^{(t)}, N_d^{G,t}(v) \mid i = 1, 2, \cdots, T\}$
17: **END_FORWARD**
18:
19: **BACKWARD**($\{\mathcal{M}^{(t)}, N_d^{G,t}(v) \mid i = 1, 2, \cdots, T\}$)
20: bulid $L_{\mathcal{M}}^{(t+1)}$ and $L_{DE}^{(t)}$ from $\mathcal{M}^{(t)}, N_d^{G,t}(v)$ by Eq. (5, 10).
21: $L^t \leftarrow \lambda_1 L_{DE}^{(t)} + (1 - \lambda_1) L_{\mathcal{M}}^{(t+1)}$
22: bulid $L_{total}$ by Eq. (15).
23: **return** $\frac{\partial L_{total}}{\partial \mathrm{w}}$
24: **END_FORWARD**

---

first conducts the unidirectional cross-propagation mechanism in order to derive the predicted matching matrix $\mathcal{M}^{(t)}$ at $t$th layer and the cross-information matrix $\mathbf{N}_Q^{(t)}$. Step 6 to step 14 is the adaptive edge-deleting mechanism. $q^{(t),k}$ is generated by the $\mathbf{H}_Q^{(t)}$, which includes the information about the query graph $Q$ and can guide how to 'delete' extra edges. After computing $\alpha_{ij}^{G,(t),k}, \alpha_{ij}^{Q,(t),k}$ from $\mathbf{N}_Q^{(t)}, \mathbf{H}_G^{(t)}, q^{(t),k}$ by Eq. (9), we can perform the multi-head GAT combining with the residual connection (Eq. (14)) to obtain the $\mathbf{H}_Q^{(t+1)}$ and $\mathbf{H}_G^{(t+1)}$. The Algorithm 2 shows our model's training process.

**Algorithm 2** Training process of AEDNet.
1: initialize w;
2: **repeat**
3:    $\{\mathcal{M}^{(t)}, N_d^{G,t}(v) \mid i = 1, 2, \cdots, T\}$ = **FORWARD**($G, Q$; w);
4:    $\frac{\partial L_{total}}{\partial w}$ = **BACKWARD**($\{\mathcal{M}^{(t)}, N_d^{G,t}(v) \mid i = 1, 2, \cdots, T\}$);
5:    w = w − $\lambda \cdot \frac{\partial L_{total}}{\partial w}$;
6: **until** (a stopping criterion)
7: **return** w

## 5. Experiments

We evaluate our model AEDNet against both SOTA learning-based approaches and exact methods on subgraph matching task in order to address the following questions: **Q1**: How accurate (effective) and fast (efficient) is AEDNet compared with both SOTA learning-based approaches and exact methods? **Q2**: How effective is the unidirectional cross-propagation mechanism and the sample-wise adaptive edge-deleting mechanism used in the AEDNet? **Q3**: How well does the proposed AEDNet adapt to the situations with noise and imbalanced graph sizes? **Q4**: Is the proposed AEDNet robust for different hyper-parameters?

### 5.1. Datasets

To assess the performance of AEDNet in identifying matching relationships between graph pairs in different domains ranging from synthetic, bioinformatics, small molecules to social networks, we use six open graph datasets, SYNTHETIC [44], COX2 [45], DD [46], PROTEINS_full [47], PPI [48] and IMDB-BINARY [49]. We take each original graph in the dataset and a connected subgraph randomly extracted from the former as a sample pair for each dataset. We then use VF2 [10] to compute the ground-truth matching matrix.

Each sample in the original dataset is one single graph. For this study, we need graph pairs with each pair including a data graph and a query graph. To construct graph pairs using the original single graph from the datasets, first we randomly select a graph from one original dataset as the data graph $G$, then randomly select a connected subgraph from this data graph $G$ as the query graph $Q$. Finally, we use the exact algorithm VF2 to calculate the ground-truth matching matrix. This operation is repeated many times to form the processed dataset. Thus, the graph pairs in the test set are unseen to the model in the training stage. Algorithm 3 de-

**Algorithm 3** generate dataset.
**Input**: $d$, original dataset. $S$, scope of size of query graph. $N$, the number of generated samples.
**Output**: $N$ samples (graph pairs).
1: $n \leftarrow 0$.
2: **while** $n < N$ **do**
3:    randomly select a graph from $d$ as $G$.
4:    randomly select $s$ from scope $S$ as the size of $Q$.
5:    randomly extract a subgraph of size $s$ from $G$ as $Q$.
6:    compute matching matix $M$ by VF2.
7:    save one sample ($G, Q, M$).
8: **end while**
9: **return** $N$ samples (graph pairs)

scribes this process. The statistics for each dataset can be found in Table 1.

**Table 1**
Summary of datasets. avg.|G| and avg.|Q| are average size of data graphs and query graphs. CU,NU and none represent categorical feature, numerical feature and none-feature respectively. The categorical feature represents the class information and the numerical feature includes many attributes.

|         | SYNTHETIC | COX2  | DD     | PROTEINS | PPI  | IMDB  |
|---------|-----------|-------|--------|----------|------|-------|
| avg.|G| | 100       | 41.22 | 284.32 | 39.06    | 2372 | 19.77 |
| avg.|Q| | 22.5      | 15    | 35     | 10       | 75   | 7.5   |
| feature | CU        | CU    | CU     | NU       | NU   | none  |

**Table 2**
Final hyper-parameter selection. $d$, $K$, $L$ are the dimension of hidden vector, the number of attention head and the number of layers respectively.

|              | SYNTHETIC   | COX2       | DD         |
|--------------|-------------|------------|------------|
| ($L, K, d$)  | (4, 8, 128) | (7, 8, 64) | (5, 6, 64) |
|              | PROTEINS    | PPI        | IMDB       |
| ($L, K, d$)  | (3, 8, 64)  | (3, 8, 128)| (3, 8, 32) |

### 5.2. Baseline methods

We first consider two learning-based subgraph matching methods, namely Sub-GMN [27] and NeuralMatch [28]. For graph matching method, we use four SOTA learning-based graph matching methods, including NGMN [25], GMNN [24], RDGCN [23], PCA-GM [21]. Exact methods VF2 [10] and VF3 (first solution) are used as baselines for the perspective of runtime.

### 5.3. Parameters setting

All details of the hyper-parameter search can be found in the Table 2. We conduct all the experiments on a single machine with an Intel Xeon 4114 CPU and one Nvidia Titan GPU. As for training, we use the Adam algorithm for optimization [50] and fix the initial learning rate to 0.001. The proposed model is trained on the training set for about 100 epochs, and checkpoints are saved for each epoch to select the best checkpoints on the evaluation set.

In our experiments, we set the dimension of hidden vector $d$ and the number of attention head $K$ of each layer to be identical for convenience. For the balance of two losses defined in Eq. (15), we set $\lambda_1 = 0.5$, $\lambda_2 = 0.2$. Also, we set identical pooling operation in Eq. (8). we perform hyper-parameter search for the dimension of hidden vector $d$, the number of attention head $K$ and the number of AEDNet layers $L$, where the searching scopes of $d$, $K$ and $L$ are in {64, 128, 256}, {1, 2, 4, 6, 8, 10} and {3, 4, 5, 6, 7, 8} respectively. The Table 2 contains the final hyper-parameter selection.

### 5.4. Evaluation metrics

We use the F1-score of the top-1 predicted subgraph to evaluate the methods due to the imbalance of sizes between the data graph and the query graph (detailed in Section 2). We further use the running time to evaluate the efficiency. Here, we select the top-1 predicted subgraph by taking the subgraph corresponding to the maximum value of each row of the predicted matching matrix $\mathcal{M}^{(T)}$. Other learning-based methods extract the top-1 predicted subgraph in the same way.

- **F1-score**
$$F_1 = \frac{2 \cdot P \cdot R}{P + R} \tag{16}$$

where $P$ is the precision representing the ratio of the number of correctly discovered node matches over the number of all discovered node matches, $R$ is the recall representing the ratio of the correctly discovered node matches over all correct node matches.

**Table 3**
The F1-score of six baselines and the proposed model on six datasets. SYN represents SYNTHETIC dataset.

|            | SYN   | COX2  | DD    | PROTEINS | PPI   | IMDB  |
|------------|-------|-------|-------|----------|-------|-------|
| NGMN       | 0.866 | 0.834 | 0.869 | 0.843    | 0.843 | 0.933 |
| GMNN       | 0.855 | 0.868 | 0.876 | 0.898    | 0.874 | 0.941 |
| RDGCN      | 0.876 | 0.846 | 0.879 | 0.883    | 0.843 | 0.930 |
| PCA-GM     | 0.921 | 0.911 | 0.899 | 0.877    | 0.892 | 0.951 |
| NeuralMatch| 0.933 | 0.919 | 0.896 | 0.888    | 0.765 | 0.949 |
| Sub-GMN    | 0.920 | 0.925 | 0.912 | 0.870    | 0.902 | 0.954 |
| AEDNet w/o C | 0.974 | 0.960 | 0.949 | 0.914  | 0.965 | 0.980 |
| AEDNet w/o D | 0.959 | 0.962 | 0.930 | 0.933  | 0.949 | 0.973 |
| AEDNet     | **0.986** | **0.979** | **0.958** | **0.949** | **0.971** | **0.980** |

**Table 4**
Average running time (seconds).

|                   | SYNTHETIC | DD    | PPI   |
|-------------------|-----------|-------|-------|
| VF2               | 0.58      | 64    | 226   |
| VF3(first solution)| 0.0005   | 0.002 | 5.82  |
| AEDNet            | 0.004     | 0.005 | 0.027 |

**Table 5**
The F1-score under different hyper-parameter.

| $(L, K, L)\backslash F1$ | COX2  | DD    | PPI   |
|--------------------------|-------|-------|-------|
| (3, 8, 128)              | 0.976 | 0.949 | **0.971** |
| (3, 6, 64)               | 0.971 | 0.944 | 0.962 |
| (5, 8, 128)              | 0.973 | 0.949 | 0.970 |
| (5, 6, 64)               | 0.971 | **0.958** | 0.965 |
| (7, 8, 128)              | 0.971 | 0.951 | 0.965 |
| (7, 6, 64)               | **0.979** | 0.950 | 0.966 |

- **Running Time** we use the running time to evaluate the efficiency of models.

### 5.5. Effectiveness and efficiency

From Table 3, we can find that our proposed model achieves SOTA results on all six data sets. It is worth noting that our proposed method still achieves an F1-score of 0.98 on the IMDB-BINARY whose graphs do not have node features, showing that our model can predict matching relationships based on structural information only. From Table 4, We found that the running time of AEDNet is faster than the exact algorithm VF3 [9] on larger graphs and VF3 is efficient on small graphs. The results show that despite being executed by python, our proposed end-to-end learning-based method is much more efficient on larger graphs. The exact methods provide 100% accuracy, yet they require more time on larger graphs.

From Table 5, different settings of hyper-parameters were tested on three datasets. We find that the proposed model performs well under different settings, the F1 scores wave within 0.014, indicating that our model is robust.

### 5.6. Ablation study

To address **Q2**, we conduct two ablation experiments. Under the AEDNet w/o C case, we exclude the unidirectional cross-propagation mechanism, meaning we do not use cross information from the data graph to replace node-level embeddings of the query graph in the previous layer. Under the AEDNet w/o D case, we remove the sample-wise adaptive edge-deleting mechanism by setting $\lambda_1 = 0$. From Table 3, we can see that the F1-score of AEDNet w/o C is 0.013 lower than AEDNet on average, and the F1-score of AEDNet w/o D is 0.019 lower than AEDNet on average, which shows that the unidirectional cross-propagation mechanism and the adaptive edge-deleting mechanism are both effective. It may be due to more similar node-level embeddings that make the attention weights more similar.

### 5.7. Analysis of effect of both noise and imbalance

Since the node feature of graphs in PROTEINS_full and PPI are numerical, we add Gaussian noise with a mean of 0 and standard deviation of 0.0625, 0.125 and 0.25 to the features to analyse the effect of noise to the proposed model. From Table 6, we can find that our proposed model still has high F1-scores under different standard deviations. The performance of our model on PROTEINS_full is better than that on PPI, mainly because the size of data graphs in PPI is larger than PROTEINS_full. Another reason could be that the magnitude of the original features of graphs in PPI is small, and a small disturbance to the features could cause heavy influence.

We measure the relationship between the size imbalance in each sample (a pair of graphs) and the F1-score on PPI dataset. The size of the graph in PPI is the largest. In Fig. 4, the x-axis is the ratio of |Q| over |G| indicating the degree of imbalance, and the y-axis is F1-score. We find that the more unbalanced the sizes of data graphs and query graphs, the lower the F1-score. On the other hand, we also find that the F1-scores on most samples are above 0.96, and the smallest F1-score is 0.92 obtained on the PPI test set, indicating that the proposed model handles node imbalance between the data graph and the query graph well.

## 6. Our limitation

In this paper, the sizes of graphs in the data set are 20 times the sizes used in the existing learning-based subgraph matching method. However, compared with some learning-free exact/approximation algorithms, such as Carletti et al. [15], Liu et al. [16], Tian et al. [17], Tian and Patel [51], the sizes of the graphs used in this study are still relatively small. For example, the size of the evaluated graphs can reach 10,000 in SOTA exact method VF3 [15]. Compared to this, the data graph used in this study is smaller and sparser. This limitation mainly comes from an essential defect of the graph neural network, called over-smoothing [29]. Graph neural networks stack multiple layers, resulting in node-level embeddings that tend to be consistent and indistinguishable from each other. However, shallow Graph neural networks have smaller receptive fields and thus can not learn global information well, especially for huge graphs.

**Table 6**
The F1-score on different standard deviation. std is the standard deviation.

| std            | 0.0625 | 0.125 | 0.25  |
|----------------|--------|-------|-------|
| PROTEINS_full  | 0.942  | 0.935 | 0.927 |
| PPI            | 0.933  | 0.904 | 0.889 |

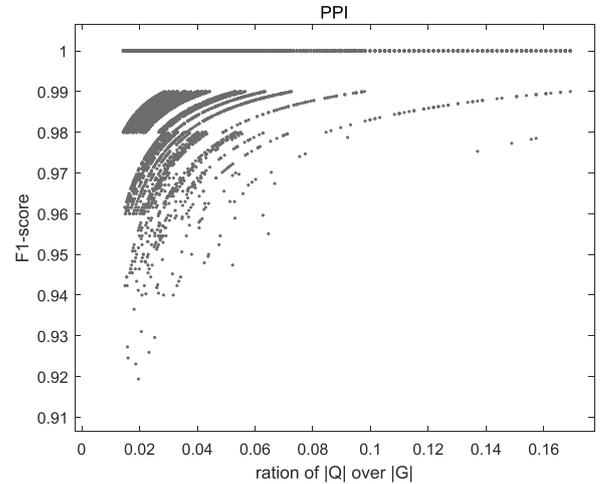

**Fig. 4.** The effect of graph size imbalance. The x-axis is the ratio of |Q| over |G|, and the y-axis is F1-score.

In this study, the algorithm used to generate all experimental graph pairs creates random query graphs from the data graphs to form graph pairs. The data graphs in our experiment always contain the query graphs. This enables us to verify the effectiveness of our proposed model on induced subgraph isomorphism. For the situation when the query graph is not contained in the data graph, the problem becomes the maximum common subgraph (MCS) isomorphism problem. Due to the limitation of our experimental dataset, we cannot verify if our method can also be applied to MCS problems. Theoretically, our method can be applied to MCS problem as well, as our method is not an exact method and it returns a set of nodes and their correspondence to the related nodes of the query graph.

## 7. Conclusion

In this paper, we proposed an AEDNet for the subgraph matching problem. It incorporates two novel mechanisms to ensure that the matched nodes' features and adjacency structures are similar. Extensive experiments on six datasets show that AEDNet outperforms six learning-based SOTA methods and is robust. Compared with other learning-based subgraph matching methods, our proposed method works effectively and efficiently on large graphs with sizes 20 times bigger. Our experiment shows that the proposed method is efficient compared with the exact method. In terms of generalization, our proposed model can accept new graphs during the inferencing that are not in the training set. Solving the subgraph matching with huge graphs is a future direction of learning-based methods.

## Data availability

AEDNet: Adaptive Edge-Deleting Network For Subgraph Matching.

## Source code availability

The source code of this paper is available at https://github.com/zixun-lan/AEDNet-Adaptive-Edge-Deleting-Network-For-Subgraph-Matching.

## Acknowledgment

This study was supported in part by the XJTLU laboratory for intelligent computation and financial technology through XJTLU Key Programme Special Fund (KSF-21) and Research Enhancement Fund of XJTLU (REF-19-01-04).